\documentclass[usenatbib]{mn2e}
\usepackage{verbatim,graphicx, amsmath, deluxetable}

\def \apjl {ApJL}
\def \apj {ApJ}
\def \aj {AJ}
\def \aap {A\&A}
\def \mnras {MNRAS}
\def \pasp {PASP}
\def \apjs {ApJS}
\def \nat {Nature}

\title[Features in Broadband Eclipse Spectra]{Features in the Broadband Eclipse Spectra of Exoplanets: Signal or Noise?}
\author[C. J. Hansen, J. C. Schwartz \& N. B. Cowan]{Christopher J. Hansen$^1$, Joel C. Schwartz$^1$  and Nicolas B. Cowan$^1$\thanks{E-mail:
ncowan@amherst.edu (NBC)}\\
$^{1}$Center for Interdisciplinary Exploration and Research in Astrophysics (CIERA), Department of Physics \& Astronomy,\\ Northwestern University, 2145 Sheridan Road, Evanston, IL, 60208, USA}
\begin{document}

\date{}

\pagerange{\pageref{firstpage}--\pageref{lastpage}} \pubyear{2002}

\maketitle

\label{firstpage}

\begin{abstract}
A planet's emission spectrum contains information about atmospheric composition and structure. We compare the Bayesian Information Criterion (BIC) of blackbody fits and idealized spectral retrieval fits for the 44 planets with published eclipse measurements in multiple thermal wavebands, mostly obtained with the Spitzer Space Telescope. The evidence for spectral features depends on eclipse depth uncertainties. \emph{Spitzer} has proven capable of eclipse precisions better than $10^{-4}$ when multiple eclipses are analyzed simultaneously, but this feat has only been performed four times.  It is harder to self-calibrate photometry when a single occultation is reduced and analyzed in isolation; we find that such measurements have not passed the test of repeatability.  Single-eclipse measurements either have an uncertainty floor of $5\times10^{-4}$, or their uncertainties have been underestimated by a factor of 3. If one adopts these empirical uncertainties for single-eclipse measurements, then the evidence for molecular features all but disappears: blackbodies have better BIC than spectral retrieval for all planets, save HD~189733b, and the few planets poorly fit by blackbodies are also poorly fit by self-consistent radiative transfer models. This suggests that the features in extant broadband emission spectra are due to astrophysical and instrumental noise rather than molecular bands.  Claims of stratospheric inversions, disequilibrium chemistry, and high C/O ratios based solely on photometry are premature. We recommend that observers be cautious of error estimates from self-calibration of small data sets, and that modelers compare the evidence for spectral models to that of simpler models such as blackbodies.
\end{abstract}


\section{Introduction}
An exoplanet on an edge-on orbit periodically passes behind its host star.  The decrement in thermal flux that occurs during such an eclipse is a measure of the dayside brightness temperature of the planet.  The brightness temperature of a planet varies with wavelength, primarily because of the atmosphere's wavelength-dependent opacity and vertical temperature profile \citep[e.g.,][]{Deming_2005, Seager_2005, Barman_2005, Burrows_2007, Burrows_2008, Fortney_2008, Knutson_2008, Desert_2009}. If different wavelengths probe the same atmospheric layer (e.g., a cloud deck) then the planet will appear to have a blackbody spectrum. In the absence of clouds, a planet may still have a blackbody spectrum if the atmospheric layers probed are isothermal.  Indeed, the emission spectra of some planets are reported to be featureless: e.g., TrES-2 \citep{ODonovan_2010}, TrES-3 \citep{Fressin_2010}, WASP-18b \citep{Nymeyer_2011}, and WASP-12b \citep{Crossfield_2012}.

In principle, the detection of molecular bands in the infrared emission spectrum of a planet enables the retrieval of greenhouse gas abundances and the vertical temperature profile of the planet \citep[e.g.,][]{Madhusudhan_2009b, Madhusudhan_2011, Lee_2012, Line_2012}.  Spectral resolution is critical to such retrieval exercises because a high-resolution emission spectrum is more likely to deviate significantly from a blackbody, and renders the retrieval problem well-constrained.  This bodes well for current and future efforts to perform \emph{bona fide} emission spectroscopy.  So far, however, the vast majority of exoplanet emission measurements have been broadband eclipse photometry.  

A typical retrieval model uses a dozen parameters to describe the atmospheric composition and vertical temperature profile, while a typical hot Jupiter has only been observed in 2--4 thermal broadbands.  Even for the few planets with 6 or 7 thermal measurements, the photometric retrieval problem is under-constrained.  

A widely noted consequence of the parameter--data mismatch is that exact atmospheric properties cannot be uniquely determined, making color-color and color-magnitude diagrams more realistic approaches to atmospheric classification \citep{Baskin_2013, Beatty_2014, Triaud_2014}. 

The less-discussed aspects of under-constrained retrieval are that (1) there is no way to reject erroneous measurements, and (2) the estimated uncertainties on eclipse depths directly affect the uncertainties on atmospheric parameters.  This is in stark contrast to over-constrained problems such as fitting an occultation model to time-series data, for which it is customary to perform outlier rejection (e.g., $\sigma$-clipping), and for which the photometric uncertainties are estimated in the process of fitting a model to the data, rather than trusting the output of aperture photometry routines.  

Nonetheless, many exoplanet discoveries have been based on broadband emission spectra: a temperature inversion in the atmosphere of HD~209458b was inferred from 4 broadband eclipse depths \citep{Knutson_2008}, disequilibrium chemistry was invoked to explain the 6-band emission spectrum of GJ~436b \citep{Stevenson_2010}, and high atmospheric C/O was discovered based on 7 broadband eclipses of WASP-12b \citep{Madhusudhan_2011}.  These successes have led to classifying planets based on temperature inversions \citep[using 2 broadbands per planet;][]{Knutson_2010} and C/O ratio \citep[using $\ge 4$ bands per planet;][]{Madhusudhan_2012}.

Temperature inversions and non-solar chemistry have since been disputed for each of the exemplar planets due to the re-reduction of existing data \citep{Beaulieu_2011} acquisition of new data at the same wavelength \citep{Cowan_2012, Zellem_2014} or acquisition of new data at different wavelengths \citep{Crossfield_2012}.  Such challenges are not unique to eclipse \emph{photometry}: the featureless day-side emission spectrum of HD~189733b \citep{Grillmair_2007} exhibited an absorption feature at a later epoch \citep{Grillmair_2008},\footnote{Although this was interpreted as evidence of planetary variability, that hypothesis is inconsistent with the more extensive monitoring campaign of \cite{Agol_2010}.} and line emission from the dayside of HD~189733b \citep{Swain_2010} has been disputed by \cite{Mandell_2011}. 

Nor are issues of repeatability limited to superior conjunction: the first thermal phase measurements of an exoplanet \citep{Harrington_2006} were later found to be off by $80^\circ$ in phase and more than a factor of 2 in amplitude \citep{Crossfield_2010}; the first half of the thermal phase measurements of \cite{Knutson_2007b} were later found to be corrupted by detector systematics \citep{Agol_2010}. 
 
The situation is similar for transit spectroscopy, where initial claims of molecular absorption bands \citep{Tinetti_2007, Swain_2008, Tinetti_2010} were disputed on the basis of data reduction, error estimation, and astrophysical variability \citep{Ehrenreich_2007, Desert_2009, Gibson_2011, Desert_2011c, Crouzet_2012}.

Indeed, \cite{Burrows_2014} offers a sobering review of the exoplanet atmospheric characterization field, speculating that many of the extraordinary claims of the past decade may be overturned by better data. 

In this article we attempt to reconcile Burrows' pessimistic view with the growing body of papers making statements about planetary atmospheres based on a handful of eclipse measurements.  Instead of focusing on a single planet, we perform a holistic analysis of all transiting planets with multiple eclipse measurements.  We consider only broadband measurements (for which it is easy to quantify the number of independent observational constraints) of eclipse depths (which are unaffected by star spots). Our approach is to compare the goodness-of-fit and evidence for three classes of models: blackbodies, self-consistent radiative transfer, and spectral retrieval.  Since the disputes over atmospheric properties have often revolved around the reliability of eclipse depths, we empirically estimate the accuracy of broadband eclipse measurements. Notably, the dominant ``signal'' in space-based eclipse photometry is usually the detector sensitivity, which must be modeled using the very same observations of the science target.  

Future observations of transiting planets with the James Webb Space Telescope are likely to resolve many of the current scientific disputes about the nature of hot Jupiter atmospheres. Attempts to push the observatory to smaller and cooler planets, however, will still rely on self-calibration; error estimation and repeatability will therefore remain critical issues.  

\section{Broadband Eclipse Spectra}
A search on exoplanet.org \citep{Wright_2011} combined with a careful literature review yields 44 exoplanets with published photometric eclipse measurements in at least two thermal wavelengths ($\lambda>1$~$\mu$m), summarized in Table~\ref{thermal_eclipses}. In most cases, only a single occultation has been measured at each wavelength. Bolded numbers signify measurements based on more data: multiple eclipses and/or an eclipse embedded in phase variations. 

\begin{table}
\caption{Planets with at least 2 thermal eclipse measurements \label{thermal_eclipses}}
\begin{tabular}{ll}
\hline
Planet & Wavelengths ($\mu$m) \\ 
\hline
CoRoT-1b & 1.65, 2.15, 3.6, 4.5\\
CoRoT-2b& 2.15, 3.6, 4.5, 8.0 \\
GJ~436b& 3.6, 4.5, 5.8, \textbf{8.0}, 16.0, 24.0\\
HAT-P-1b& 3.6, 4.5, 5.8, 8.0\\
HAT-P-2b& 3.6, 4.5, 5.8, 8.0\\
HAT-P-3b& 3.6, 4.5\\
HAT-P-4b& 3.6, 4.5\\
HAT-P-6b& 3.6, 4.5\\
HAT-P-7b& 3.6, 4.5, 5.8, 8.0\\
HAT-P-8b& 3.6, 4.5\\
HAT-P-12b& 3.6, 4.5\\
HAT-P-23b& 2.15, 3.6, 4.5\\
HD~149026b& 3.6, 4.5, 5.8, \textbf{8.0}\\
HD~189733b& 2.15, \textbf{3.6}, \textbf{4.5}, 5.8, \textbf{8.0}, 16.0, 24.0\\
HD~209458b& 2.15, 3.6, \textbf{4.5}, 5.8, 8.0, \textbf{24.0}\\
KELT-1b& 3.6, 4.5\\
Kepler~5b& 3.6, 4.5\\
Kepler-6b& 3.6, 4.5\\
Kepler-12b& 3.6, 4.5\\
Kepler-13Ab& 2.15, 3.6, 4.5\\
Kepler-17b& 3.6, 4.5\\
TrES-1b& 3.6, 4.5, 8.0\\
TrES-2b& 2.15, 3.6, 4.5, 5.8, 8.0\\
TrES-3b& 1.25, 2.15, 3.6, 4.5, 5.8, 8.0\\
TrES-4b& 3.6, 4.5, 5.8, 8.0\\
WASP-1b& 3.6, 4.5, 5.8, 8.0\\
WASP-2b& 3.6, 4.5, 5.8, 8.0\\
WASP-3b& 3.6, 4.5, 8.0\\
WASP-4b& 2.15, 3.6, 4.5\\
WASP-5b& \textbf{1.25}, \textbf{1.65}, \textbf{2.15}, 3.6, 4.5\\
WASP-8b& 3.6, 4.5, 8.0\\
WASP-12b& 1.25, 1.65, 2.15, \textbf{3.6}, \textbf{4.5}, 5.8, 8.0\\
WASP-14b& 3.6, 4.5\\
WASP-17b& 4.5, 8.0\\
WASP-18b& \textbf{3.6}, \textbf{4.5}, 5.8, 8.0\\
WASP-19b& 1.65, 3.6, 4.5, 5.8, 8.0\\
WASP-24b& 3.6, 4.5\\
WASP-33b& 2.15, 3.6, 4.5\\
WASP-43b& 3.6, 4.5\\
WASP-48b& 1.65, 2.15, 3.6, 4.5\\
XO-1b& 3.6, 4.5, 5.8, 8.0\\
XO-2b& 3.6, 4.5, 5.8, 8.0\\
XO-3b& 3.6, \textbf{4.5}, 5.8, 8.0\\
XO-4b&3.6, 4.5\\	
\hline
\end{tabular}\\
\citep[][]{Agol_2010, Alonso_2010, Anderson_2010a, Anderson_2011, Barnes_2007, Baskin_2013, Beatty_2014, Beaulieu_2011, Beerer_2011, Blecic_2013, Campo_2011, Charbonneau_2005, Charbonneau_2008, Chen_2014, Christiansen_2009, Cowan_2012, Croll_2010a, Croll_2010b, Croll_2011, Crossfield_2012, Cubillos_2013, Deming_2005, Deming_2006, Deming_2007, Deming_2011, Demory_2007, Desert_2011a, Desert_2011b, Fortney_2011, Fressin_2009, Gillon_2009a, Gillon_2009b, Knutson_2007a, Knutson_2007b, Knutson_2008, Knutson_2009a, Knutson_2009b, Knutson_2009c, Knutson_2012,  Lewis_2013, Lopez-Morales_2010, Machalek_2008, Machalek_2009, Machalek_2010, deMooij_2013, Nymeyer_2011, ODonovan_2010, ORourke_2014, Richardson_2003, Rogers_2009, Rostron_2014, Shporer_2014, Smith_2012, Stevenson_2010, Stevenson_2012, Stevenson_2014, Todorov_2009, Todorov_2012, Todorov_2013, Wheatley_2010, Wong_2014, Zellem_2014, Zhou_2013}.
\end{table}

Since we are merely concerned with the emergent spectra of the bodies at superior conjunction, it is immaterial if a planet has an eccentric orbit (GJ 436b, HAT-P-2b, WASP-8b, WASP-14b, XO-3b) or is a highly-irradiated brown dwarf (KELT-1b). The majority of these observations---in particular, all those longward of 3~$\mu$m--- were made with the Spitzer Space Telescope \citep[][]{Werner_2004}. In cases where multiple values have been published, we adopt the most recent.

We fit a blackbody spectrum to the eclipse depths for each planet using the published transit depth and stellar effective temperature. We assume symmetric, Gaussian, error bars for the eclipse depths; in the few cases were asymmetric error bars were published, we take the mean of the upper and lower error bars.  The transit depth and stellar effective temperature have associated uncertainties that tend to have a gray impact on the planet's spectrum and hence we neglect them in the current analysis. 

In the interest of simplicity, we ignore the detector spectral response functions and instead compute the Plank function at the central wavelength of each photometric observation. Moreover, by using the stellar effective temperature rather than a detailed stellar model, we are treating the star as a blackbody.  These assumptions are reasonable for broadband measurements in the infrared.

\section{The Significance of Spectral Features}
A spectral retrieval model can provide a better fit to observations than a blackbody, because it has roughly a dozen free parameters, rather than one.  A self-consistent radiative transfer model lies somewhere in between, with a few variables. In order to compare the evidence for these models, we use the Bayesian Information Criterion \citep[BIC;][]{Schwarz_1978}.  BIC is a simple way to compare the evidence for models with different numbers of parameters: $BIC = \chi^2 + k\ln{N}$, where $\chi^2$ is the usual badness-of-fit, $k$ is the number of free parameters and $N$ is the number of data. It is similar in spirit to the reduced $\chi^2$ in that it penalizes models with many parameters, but it remains well-defined when there are fewer data than there are parameters, as is the case for current photometric eclipse retrieval.  The Akaike Information Criterion \citep[AIC;][]{Akaike_1974} penalizes complex models even more than the BIC for $N<7.4$, i.e., for all of the planets considered here.  Moreover, \cite{Chen_2008} note that both BIC and AIC tend to be biased in favor of complex models in the small-$N$, large-$k$ regime.  In short, our use of the BIC gives models with many free parameters the benefit of the doubt.
  
\begin{figure}
\centering
\includegraphics[width=0.5\textwidth]{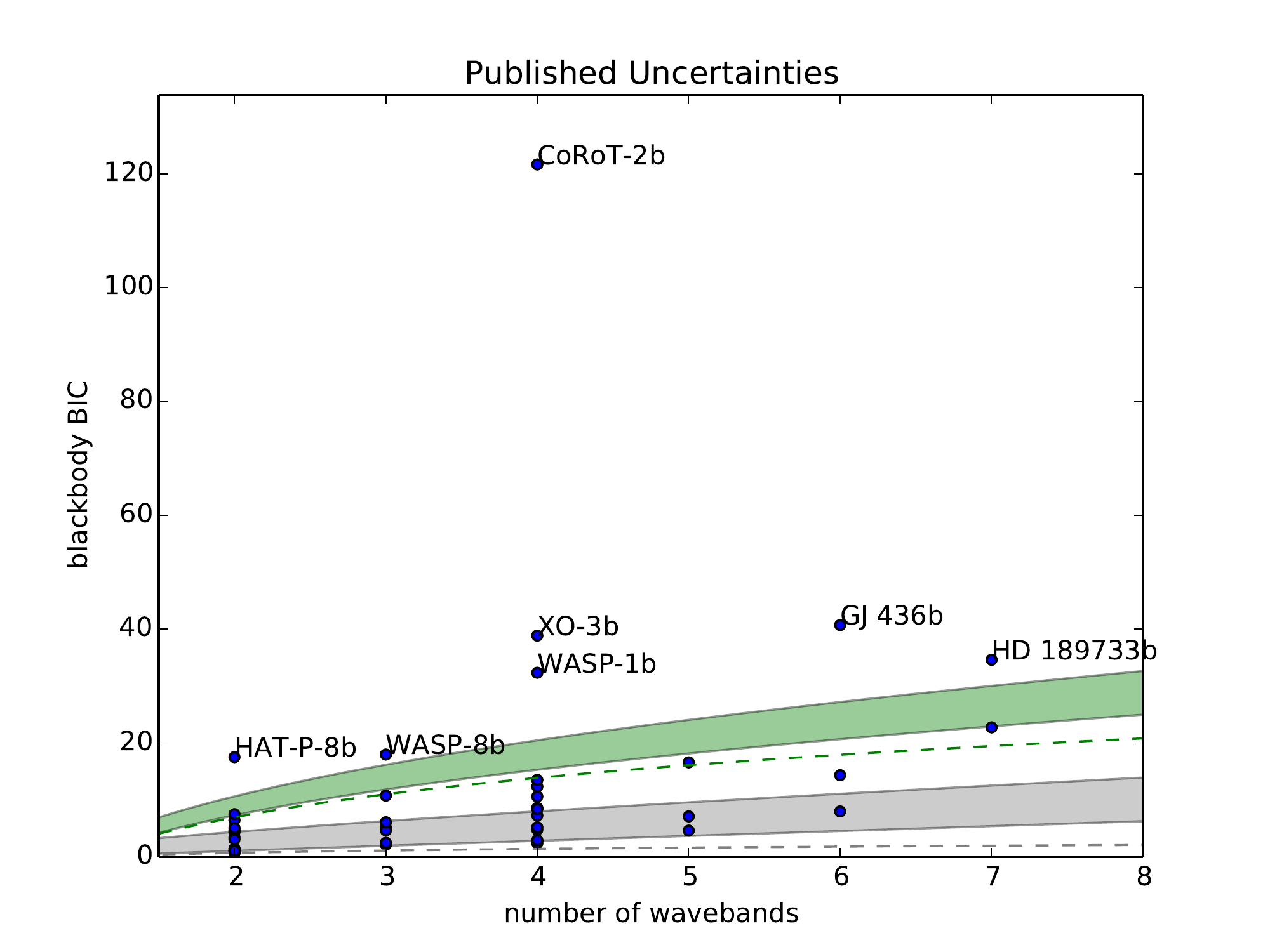} 
\caption{The Bayesian Information Criterion (BIC) of a blackbody fit is plotted against the number of thermal wavebands for which photometric eclipse measurements have been obtained; each dot represents one of the 44 transiting planets in our sample. The dashed gray line is a perfect fit to a blackbody ($\chi_{\rm BB}^2=0$), while the gray region denotes a good fit ($\chi_{\rm BB}^2/N\approx1$ with 68.3\% confidence interval). Planets that lie well above the gray region are poorly fit by a blackbody; the vertical distance above the gray indicates the strength of broadband features in that planet's emission spectrum. Green denotes the quality of a hypothetical spectral retrieval fit: the dashed line is a perfect fit ($\chi_{\rm SR}^2=0$), while the green region is a good fit ($\chi_{\rm SR}^2/N\approx1$ with 68.3\% confidence interval). Planets that lie in or above the green region may favor spectral retrieval, if published uncertainties are taken at face value.\label{blackbody_BIC}}
\end{figure}

As a baseline, we fit a blackbody and compute the BIC for each planet in our sample using the published eclipse depths and uncertainties. The only unknown is the blackbody temperature, so $k=1$ and $BIC_{\rm BB} = \chi_{\rm BB}^2 + \ln{N}$. Figure~\ref{blackbody_BIC} shows the blackbody BIC plotted against the number of wavebands available for each planet. Gray denotes the quality of a blackbody fit: the dashed gray line is a perfect fit to a blackbody ($\chi_{\rm BB}^2=0$), while the gray region denotes a good fit ($\chi_{\rm BB}^2/N\approx1$ with 68.3\% confidence interval). 

Since there are few data, the $\chi^2$ distribution is broad and asymmetrical, with a tail towards large values (the colored swaths denote the 68.3\% intervals of the $\chi^2$ distribution). Planets that lie well above the gray region are poorly fit by a blackbody, given the published uncertainties. The vertical distance above the gray indicates the strength of broadband features in that planet's emission spectrum. CoRoT-2b exhibits by far the most featured broadband emission spectrum of any transiting planet, a fact not lost on observers \citep[e.g.,][]{Cowan_2011_sptz}. 

We also consider an idealized spectral retrieval model with 10 free parameters: 6 parameters for the vertical temperature--pressure profile and 4 for molecular abundances \citep[][]{Madhusudhan_2009b}. Some recent retrieval studies have 2 additional abundance variables, for a total of 12 model parameters \citep[e.g.,][]{Stevenson_2014}, so our adoption of 10 is conservative.  Since it is under-constrained, one might expect spectral retrieval to provide perfect fits to broadband emission spectra (i.e., $\chi_{\rm SR}^2=0$).  We denote this scenario with the dashed green line in Figure~\ref{blackbody_BIC} ($BIC_{\rm SR} = 10\ln{N}$). In practice, spectral retrieval involves \emph{a priori} constrains (e.g., priors on plausible chemistry) so their fits have been in the range $\chi_{\rm SR}^2/N=0.5$--2 \citep{Madhusudhan_2010, Madhusudhan_2011, Madhusudhan_2012}. We therefore also plot a green region denoting $BIC_{\rm SR} = \chi_{\rm SR}^2+10\ln{N}$ (i.e., a spectral retrieval fit with $k=10$ and $\chi_{\rm SR}^2/N\approx1$). 

We expect that spectral retrieval would produce BIC values in the green swath.  While the derivation of BIC relies on assumptions that may not be entirely valid for spectral retrieval, planets that lie above the green region exhibit a preference for spectral retrieval as compared to a blackbody fit ($BIC_{\rm SR}<BIC_{\rm BB}$). For example, CoRoT-2b has been well fit using spectral retrieval \citep[$\chi_{\rm SR}^2/N = 0.725$;][]{Madhusudhan_2012}; if the published eclipses are taken at face value, then there is very strong evidence that spectral retrieval is a better model than a blackbody for this planet.

If published eclipse values and uncertainties are taken at face value, then many hot Jupiters lie above the gray region, indicating that they are poorly fit by blackbodies, but below the green region, implying that the poorly-fitting blackbody is favored over spectral retrieval, according to the BIC.  While one could perform spectral retrieval on these data and conceivably obtain interesting atmospheric constraints, they should be taken with a grain of salt because spectral retrieval is probably the \emph{wrong model} given the current data.

Figure~\ref{blackbody_BIC} shows seven planets with broadband emission spectra that invite a full spectral retrieval: CoRoT-2b, GJ~436b, HAT-P-8b, HD~189733b, WASP-1b, WASP-8b, and XO-3b. This list includes a few of the best/brightest transiting targets in GJ~436b, HD~189733b, and XO-3b.  Since the Poisson (photon-counting) noise is smaller for bright targets, the smaller error bars might reveal intrinsic molecular bands present in planetary emission spectra.  Alternatively, the eclipse uncertainties for bright targets may be dominated by systematic error rather than Poisson noise. Since it is notoriously difficult to estimate systematic errors \citep{Topping_1955}, it is critical to empirically evaluate the eclipse accuracy via repeated measurements \citep{Lyons_1992}.

\section{Empirical Estimate of Eclipse Uncertainties}
The instruments currently used to measure exoplanet eclipses are pushed orders of magnitude beyond their design specifications for the simple reason that transiting short-period planets were not known to exist when the instruments were designed \citep[e.g., the 2\% stability of \emph{Spitzer} IRAC;][]{Fazio_2004}. The raw photometry therefore suffers from detector systematics that are comparable to, and sometimes dwarf, the astrophysical signal of interest \citep[e.g.,][]{Charbonneau_2005, Deming_2005}.  

In what follows, we focus on \emph{Spitzer} because a) 133 of 154 published broadband thermal eclipse measurements were obtained with this telescope, b) these observations have the smallest quoted uncertainties and hence place the strongest constraints on atmospheric structure and composition, and c) these are essentially the only thermal eclipse measurements to have been repeated. 

New observing modes with \emph{Spitzer} have improved the data quality over the past decade: staring rather than dithering, only observing in a single waveband at a time, increasing the frequency of the heater cycling, and the peak-up method for keeping the target centroid on the same region of a pixel throughout long observations. Furthermore, there have been improvements in our understanding of \emph{Spitzer} systematics, especially for large data-sets, including pixel-by-pixel ramp correction \citep[][]{Knutson_2007a}, polynomial decorrelation \citep{Knutson_2008}, double-exponential ramp correction \citep[][]{Agol_2010}, Gaussian decorrelation \citep[][]{Ballard_2010}, BLISS mapping \citep[][]{Stevenson_2012}, and use of the noise pixel \citep[][]{Knutson_2012, Lewis_2013}.  It is now routine for combined detector$\times$astrophysics models to fit the data within 10--20\% of Poisson noise. 

Despite excellent fits, residuals usually exhibit red (time-correlated) noise. The wavelet-based method of \cite{Carter_2009} has been used to estimate the impact of red noise on eclipse depth uncertainties \citep[e.g., the full-orbit phase curves of HD~189733b;][]{Knutson_2012}, and Independent Component Analysis \citep{Waldmann_2012} has been used to perform blind signal de-mixing for transit spectroscopy \citep{Waldmann_2013}. Although these methods are better motivated than quick-and-dirty methods such as residual binning and residual permutation \citep{Cowan_2012}, none seem to produce accurate error bars in numerical tests: uncertainty estimates are still too small in the presence of red noise and na\"ive methods often perform best \citep[][]{Cubillos_2014}. 

In order to avoid these subtleties of error estimation we would like to fit the data so well that there is no red noise in the residuals.  This drives observers to use increasingly complex models.  It is notable that the current leading detector models for \emph{Spitzer} channels 1 \& 2 are non-parametric \citep{Ballard_2010, Knutson_2012, Stevenson_2012, Lewis_2013}. This is commonly taken to mean that they have \emph{no} free parameters, but it might be more accurate to say that they have a large, but vague, number of parameters.\footnote{The Gaussian decorrelation scheme of \cite{Knutson_2012} and \cite{Lewis_2013} has an effective number of detector parameters roughly equal to the area of the centroid range, $\Delta x \Delta y$ divided by the Gaussian smoothing area, $\sigma_x \sigma_y$.  This quantity is typically in the hundreds.} One therefore has to be wary of over-fitting, and should strive to compare models of varying complexity in a Gaussian framework.

Instead of debating the merits of detector models and uncertainty estimation schemes, we now consider the empirical accuracy of eclipse measurements.   

\subsection{Parallel Analysis of Multiple Eclipses}\label{best_case}
The ideal way to determine the uncertainty on a measurement is to repeat it: obtain many ($>2$) eclipse measurements and their standard deviation should be a robust measure of the eclipse uncertainty.  This exercise has been performed five times with \emph{Spitzer}: 6 eclipses of HD~189733b at 8~$\mu$m \citep{Agol_2010}, 11 eclipses of GJ~436b at 8~$\mu$m \citep{Knutson_2011}, 4 eclipses of 55~Cancri~e at 4.5~$\mu$m \citep{Demory_2012}, 3 eclipses of HD~209458b at 24~$\mu$m \citep{Crossfield_2012}, and 12 eclipses of XO-3b at 4.5~$\mu$m \citep{Wong_2014}.  These studies report $1\sigma$ variance of $9\times10^{-5}$, $8\times10^{-5}$, $6\times10^{-5}$, $4\times10^{-4}$, and $8\times10^{-5}$, respectively, which represent a combination of the astrophysical dayside variability of the planet, Poisson noise, and the level at which researchers could model the detector sensitivity.

\subsection{Reanalysis of Single Eclipses}\label{reanalysis}
In a few cases, the \emph{same} data have been reanalyzed and republished by different authors, and these measurements have usually differed by $<1\sigma$: HD~189733b at 16~$\mu$m \citep[][]{Deming_2006, Charbonneau_2008}, HD~149026b at 8.0~$\mu$m \citep[][]{Knutson_2009c, Stevenson_2012}, GJ~436b at 8~$\mu$m \citep[][]{Deming_2007, Demory_2007, Stevenson_2010}, and CoRoT-2b at 4.5 and 8.0~$\mu$m \citep{Gillon_2010, Deming_2011}.   

Consider, however, the reanalysis of the original \cite{Harrington_2007} 8~$\mu$m eclipse of HD~149026b by \cite{Knutson_2009b}.  The latter authors found they could reproduce the original deep eclipse measurement, as well as the new, shallow depth obtained as part of thermal phase variations: ``The diversity of eclipse depths (0.05\%--0.09\%) obtained in these fits suggests that the final result is sensitive to our specific choice of functions, fitting routines, and bad pixel trimming methods.''  

Finally, there are the secondary eclipses of GJ~436b \citep{Stevenson_2010} that were re-analyzed by \cite{Beaulieu_2011}.  The latter authors found compatible values at 5.8~$\mu$m and identical values at 8.0~$\mu$m. At 3.6~$\mu$m they found that their eclipse depth depends on the reduction scheme and details of fitting, while at 4.5~$\mu$m they also favored a non-detection, but with an uncertainty $3\times$ greater than the original authors.

\subsection{Serial Analysis of Multiple Eclipses}
For a handful of the best and brightest targets, multiple \emph{Spitzer} eclipse observations have been obtained with the same instrument and published in \emph{separate} papers. This is an important test of repeatability because it is semi-blind: the authors of the first paper did not benefit from knowing the result of subsequent observations (the latter authors, of course, had access both to the original and their new observations).  This is in contrast to the studies listed in \S\ref{best_case}, for which researchers considered the ensemble of eclipse measurements as they fine-tuned their reduction and analysis pipeline.

The results of ten semi-blind repeatability tests are listed in Table~\ref{reshoots}.  For each planet+waveband combination, we list the first published eclipse measurement based on a simple eclipse measurement, then a subsequent measurement obtained as part of thermal phase measurements or a multi-eclipse campaign. Note that for the HD~189733b 8~$\mu$m eclipse, the simple eclipse measurement \citep{Charbonneau_2008} was published \emph{after} the phase+eclipse measurement of \cite{Knutson_2007b}, but clearly the order in which we list the measurements in no way impacts the analysis below.  

For each eclipse measurement, we list the published value and uncertainty, $\sigma$.  For each pair of measurements, we list the discrepancy, $\Delta$, between the new measurement and the original. We also estimate the total published uncertainty as the quadrature sum of the first and second eclipse uncertainties: $\sigma_{\rm tot} = \sqrt{\sigma_1^2 + \sigma_2^2}$.

\begin{table*}
\caption{\emph{Spitzer} eclipse repeat observations \label{reshoots}}
\scriptsize
\begin{tabular}{lcccccccc}
\hline
Planet & $\lambda$ ($\mu$m) & Value & $\sigma$ & $\Delta$ & $\sigma_{\rm tot}$& $\sqrt{\Delta^2-\sigma_{\rm tot}^2}$ & $|\Delta|/\sigma_{\rm tot}$& Reference\\ 
\hline
GJ~436b&	8.0	&	$5.4\times 10^{-4}$	&$8.0\times 10^{-5}$&	&&&&\cite{Deming_2007}\\
&		&	$4.52\times 10^{-4}	$	&$2.7\times 10^{-5}$&		$-8.8\times 10^{-5}$&$8.4\times10^{-5}$	&$2.6\times10^{-5}$& 1.0&\cite{Knutson_2011}	\\
HD~149026b &	8.0	&	$8.4\times 10^{-4}$	&$1.0\times 10^{-4}$&&	&&&\cite{Harrington_2007}  \\		
&		&	$4.11\times 10^{-4}$	&$7.6\times 10^{-5}$&	$-4.3\times 10^{-4}$&$1.3\times10^{-4}$	&$4.1\times10^{-4}$ & 3.4 &\cite{Knutson_2009c}	\\
HD~189733b&	3.6	&	$2.56\times 10^{-3}$	&$1.4\times 10^{-4}$&	&&&&\cite{Charbonneau_2008}\\	
&		&	$1.466\times 10^{-3}$	&$4.0\times 10^{-5}$&	$-1.1\times 10^{-3}$&$1.5\times10^{-4}$	&$1.1\times10^{-3}$& 7.5&\cite{Knutson_2012}	\\
&	4.5	&	$2.14\times 10^{-3}$	&$2.0\times 10^{-4}$&	&&&&\cite{Charbonneau_2008}\\	
&		&	$1.787\times 10^{-3}$	&$3.8\times 10^{-5}$&	$-3.5\times 10^{-4}$&$2.0\times10^{-4}$	& 	$2.9\times10^{-4}$& 1.7&\cite{Knutson_2012}		\\
&	8.0 &	$3.91\times 10^{-3}$	& $2.2\times 10^{-4}$ & 	&	 & & &  \cite{Charbonneau_2008}	\\
& &	$3.381\times 10^{-3}$	&$5.5\times 10^{-5}$&	$-5.3\times10^{-4}$&	$2.3\times10^{-4}$& $4.8\times10^{-4}$& 2.3 &\cite{Knutson_2007b}	\\
HD~209458b&	4.5	&	$2.13\times 10^{-3}$	&$1.5\times 10^{-4}$&	&&&&\cite{Knutson_2008}\\	
&		&	$1.391\times 10^{-3}$	&$7.1\times 10^{-5}$&	$-7.4\times 10^{-4}$&$1.7\times10^{-4}$	&$7.2\times10^{-4}$& 4.3&\cite{Zellem_2014}	\\
&	24	&	$2.60\times 10^{-3}$	&$4.6\times 10^{-4}$&&	&&&\cite{Deming_2005}  \\		
&		&	$3.38\times 10^{-3}$	&$2.6\times 10^{-4}$&	$+7.8\times 10^{-4}$&$5.3\times10^{-4}$	&$5.7\times10^{-4}$ & 1.5 &\cite{Crossfield_2012_24}	\\
WASP-12b&	3.6	&	$3.79\times 10^{-3}$	&$1.3\times 10^{-4}$&	&&&&\cite{Campo_2011}\\
&		&	$3.3\times 10^{-3}	$	&$4.0\times 10^{-4}$&		$-4.9\times 10^{-4}$&$4.2\times10^{-4}$	 &$2.5\times10^{-4}$& 1.2&\cite{Cowan_2012}	\\
&	4.5	&	$3.82\times 10^{-3}$	&$1.9\times 10^{-4}$&	&&&&\cite{Campo_2011}\\
&		&	$3.9\times 10^{-3}	$	&$3.0\times 10^{-4}$&$+8.0\times 10^{-5}$&	$3.6\times10^{-4}$	&0&0.2&\cite{Cowan_2012}\\	
XO-3b&	4.5	&	$1.43\times 10^{-3}$	&$6.0\times 10^{-5}$&	&&&&\cite{Machalek_2010}\\
&		&	$1.580\times 10^{-3}	$	&$3.6\times 10^{-5}$&$+1.5\times 10^{-4}$&	$7\times10^{-5}$	&$1.3\times10^{-4}$ &2.1&\cite{Wong_2014}\\	
\hline
\end{tabular}
\end{table*}

Comparing the $\Delta$ and $\sigma_{\rm tot}$ columns of Table~\ref{reshoots} suggests that published eclipse uncertainties are too small: the original researchers, subsequent researchers, or both groups under-estimated the uncertainty in their measurement. Since the latter eclipse  measurements  are based on more data, we assume that they represent an accurate measurement and uncertainty, while the original measurements, based on a simple occultation, had under-estimated error bars.

Alternatively, the planets may be exhibiting weather that changes the eclipse depths from one epoch to the next, as predicted by \cite{Rauscher_2007}. Eclipse depth variability at the level of $5\times 10^{-4}$ would invalidate spectral retrieval because multi-band broadband emission spectra are constructed over a span of many planetary orbits. The weather hypothesis is ruled out in a few cases by the repeat observations discussed in \S\ref{best_case}, however. 

\subsection{Realistic Eclipse Uncertainties}
We quantify the degree to which eclipse uncertainties have been under-estimated by combining $\Delta$ and $\sigma_{\rm tot}$ to obtain an empirical estimate of systematic uncertainty, following \S2.1 of \cite{Lyons_1992}. 

In the first case, we assume there is an additional source of noise that affects single-eclipse measurements.  Physically, this might correspond to how well one can model the detector given only a few hour observation of the science target.  We estimate the magnitude of this systematic uncertainty by considering the distribution of $\sqrt{\Delta^2-\sigma_{\rm tot}}$.  In the one case where the epoch-to-epoch discrepancy, $\Delta$, was smaller than the total published uncertainty, we set this quantity to zero.

The symmetric\footnote{The $\Delta$-distribution is decidedly asymmetrical: researchers analyzing single-eclipse measurements have over-estimated the eclipse depth more often than not. Identifying the cause of this bias is beyond the scope of the current manuscript, so we limit ourself to properly estimating the empirical eclipse uncertainty.} distribution $[\sqrt{\Delta^2-\sigma_{\rm tot}}] \cup [-\sqrt{\Delta^2-\sigma_{\rm tot}}]$ has a standard deviation of $\sigma_{\rm syst} \approx 5.2\times10^{-4}$.  We adopt $\sigma_{\rm syst} = 5\times10^{-4}$ for the remainder of this paper \citep[this is somewhat greater than, but broadly consistent with, the repeatability estimate of $2\times 10^{-4}$ based on a pair of 3.6~$\mu$m transits of HD~189733b;][]{Morello_2014}. 

The second approach is to consider the distribution of $|\Delta|/\sigma_{\rm tot}$, which amounts to hypothesizing that single-eclipse uncertainties have been under-estimated by a constant factor.  For example, researchers may under-estimate the degree to which the unknown detector model impacts eclipse depth uncertainty \citep[numerical experiments have shown that most extant methods underestimate occultation error bars in the presence of correlated noise;][]{Cubillos_2014}.
The standard deviation of the symmetric distribution $[|\Delta|/\sigma_{\rm tot}] \cup [-|\Delta|/\sigma_{\rm tot}]$ is $f_{\rm syst} \approx 3.3$.  We adopt $f_{\rm syst} = 3$ in the remainder of this paper.  

\section{Broadband Spectra with Empirical Uncertainties}
To summarize the previous section, \emph{Spitzer} has proven capable of photometry better than $10^{-4}$ and many existing eclipse measurements are likely accurate at that level: specifically, those based on multiple eclipses or taken as part of longer phase measurements (the bolded numbers in Table~\ref{thermal_eclipses}). Single-epoch eclipse measurements of the best and brightest targets have \emph{not} been repeatable at this level, however.  This is unfortunate because such single-eclipse measurements represent the vast majority of the broadband emission data (the unbolded numbers in Table~\ref{thermal_eclipses}).  

\begin{figure}
\centering
\includegraphics[width=0.5\textwidth]{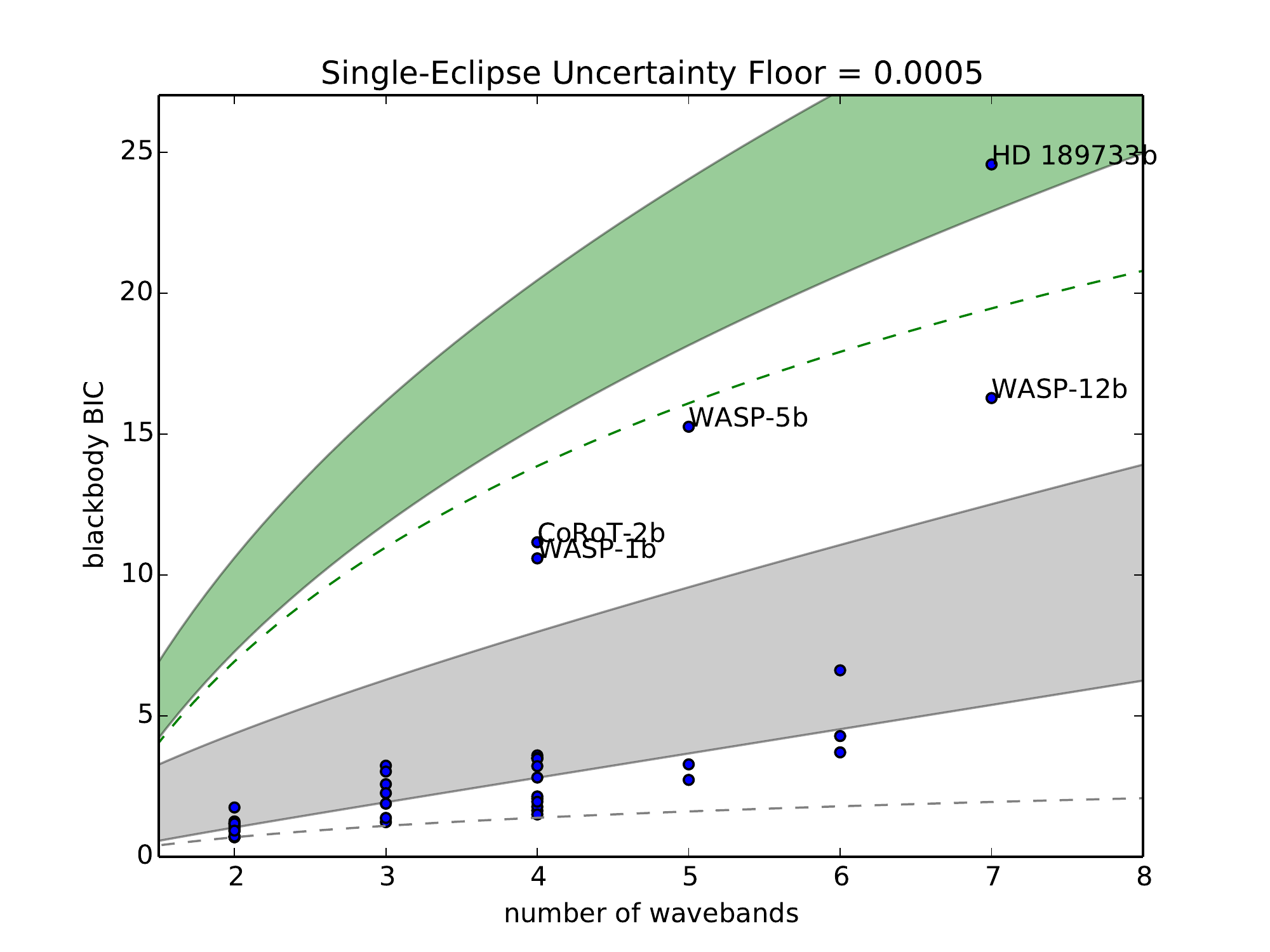}
\caption{As in Figure~\ref{blackbody_BIC}, but we add an empirical systematic error of $\sigma_{\rm syst} = 5\times 10^{-4}$ in quadrature to each simple-eclipse measurement. In this hypothesis, there is a floor to how precise an eclipse measurement can be without acquiring more data, so modern eclipse measurements are no more accurate than earlier attempts. Eclipse uncertainties based on multiple eclipse measurements, or an eclipse embedded in a phase measurement, are kept unchanged.\label{realistic_error}}
\end{figure}

Figure~\ref{realistic_error} shows the distribution of blackbody BIC vs.\ $N_{\lambda}$ in light of empirical eclipse depth uncertainties. Values based on multiple eclipse measurements, or obtained as part of phase measurements, are taken at face value.  We add a systematic uncertainty of $\sigma_{\rm syst}=5\times 10^{-4}$ in quadrature to the quoted uncertainties for all single-eclipse measurements.\footnote{If we had instead assumed that both the original and subsequent measurements were equally error-prone, then $\sigma_{\rm syst}$ and $f_{\rm syst}$ would be somewhat smaller, but they would have to be applied across the board, leaving our conclusions essentially unchanged.} We then re-fit a blackbody and recompute the BIC for each planet using these more realistic error bars.  

\begin{figure}
\centering
\includegraphics[width=0.5\textwidth]{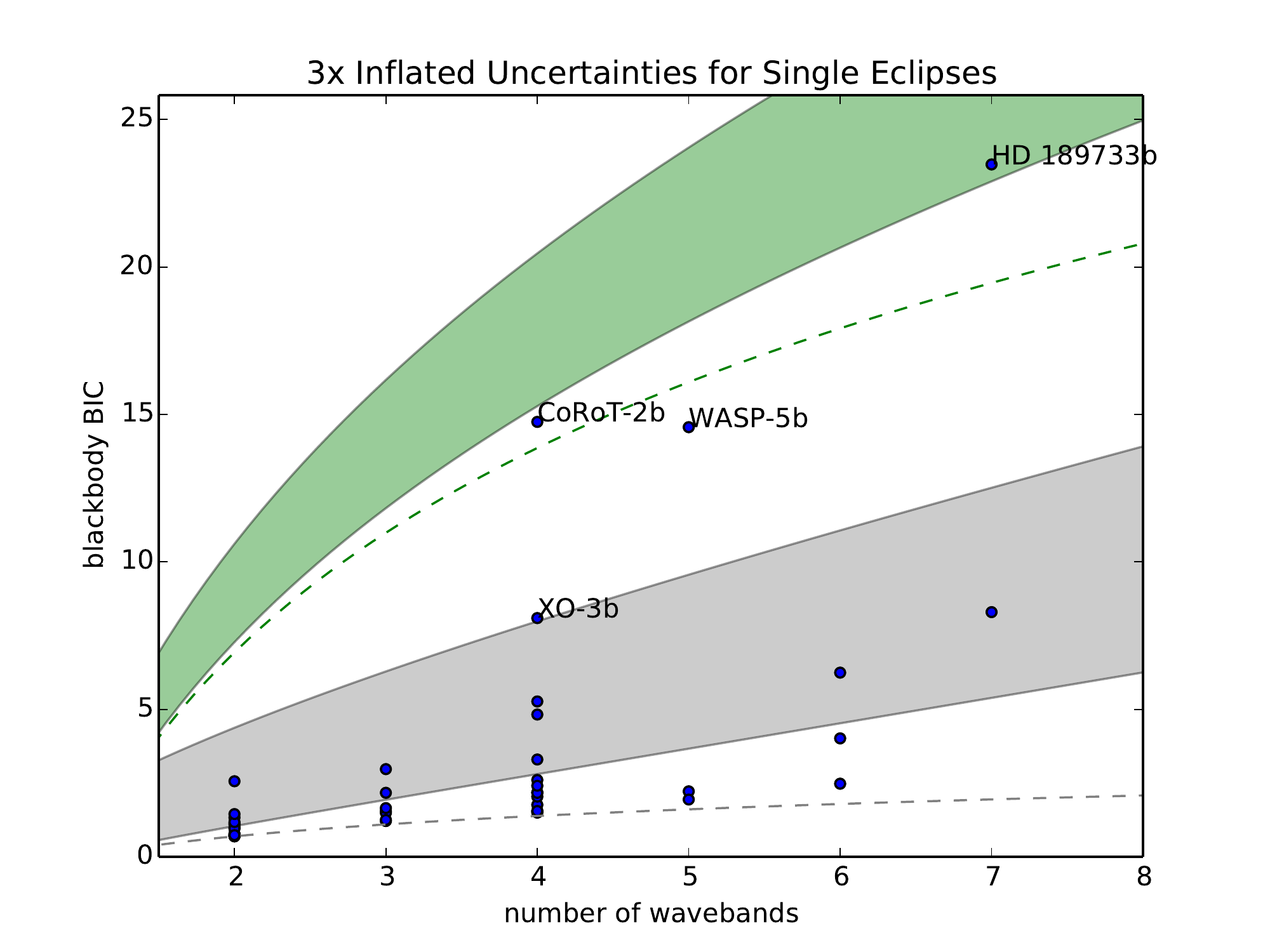}
\caption{As in Figure~\ref{blackbody_BIC}, but we inflate the published single-eclipse uncertainties by the empirical factor $f_{\rm syst} = 3$. This scenario accounts for the possibility that modern eclipse measurements, which have much smaller quoted uncertainties than the first generation of eclipses, might really be more accurate than their predecessors. Eclipse uncertainties based on multiple eclipse measurements, or an eclipse embedded in a phase measurement, are kept unchanged.\label{error_factor}}
\end{figure}

In Figure~\ref{error_factor} we inflate the published uncertainties of single-eclipse measurements by our empirically determined factor of $f_{\rm syst} = 3$. We then re-fit a blackbody and recompute the BIC for each planet using these more realistic error bars.

Under the assumption of realistic eclipse uncertainties, HD~189733b has the most featured emission spectrum and lies in the green region in both Figures~\ref{realistic_error} and \ref{error_factor}. If spectral retrieval could achieve a perfect fit, $\chi_{\rm SR}^2/N=0$, then it would be modestly favored as compared to the blackbody, according to the BIC. Obtaining such a good fit is not trivial for this planet because even our realistic noise hypothesis takes the published uncertainties at 3.6, 4.5, and 8.0~$\mu$m at face value.  

All other planets lie at/below the dashed green line, suggesting that blackbodies are favored, \emph{even if spectral retrieval provides a perfect fit to the data}.  In any case, a researcher who has gone to the trouble of running a Markov Chain Monte Carlo to perform spectral retrieval should also estimate the evidence for their model using the posterior distribution; BIC is merely a way of approximating this.  Ideally, the evidence for spectral retrieval models with different numbers of parameters could be compared using, for example, a Reversible Jump Markov Chain Monte Carlo \citep{Green_1995} or Nested Sampling \citep{Skilling_2004}.

\section{Discussion}
\subsection{The Exceptions Prove the Rule}
Given the small number statistics, we expect a broad range of $\chi^2$ values with a significant tail; the gray zone indicates the $1\sigma$ (68.3\%) interval. Nonetheless, a few short period planets lie well above the gray region in Figures~\ref{realistic_error} and \ref{error_factor}, suggesting they are poorly fit by a blackbody and hence exhibit spectral features.  These features are either the hints of molecular bands, or remaining astrophysical/detector noise. The only planets that make the cut under both the $\sigma_{\rm syst}$ and $f_{\rm syst}$ hypotheses are CoRoT-2b, HD~189733b, and WASP-5b. In order to put the poorly-fitting blackbodies in perspective, we compare them to self-consistent radiative transfer models. 

Self-consistent atmospheric radiative transfer models typically have between one and three tunable parameters: recirculation efficiency, optical opacity, and relative abundance of CO \citep[e.g.,][]{Kipping_2011, Deming_2011} and are usually tuned by eye in order to obtain a decent fit.  In what follows we will quote $\chi_{\rm RT}^2$ values from the literature (i.e., using published eclipse uncertainties).  As such, the values should be compared to the blackbody BIC values shown in Figure~\ref{blackbody_BIC}.

As noted by \cite{Deming_2011}, CoRoT-2b is so poorly fit by spectral models at 4.5~$\mu$m that a blackbody fit has a smaller $\chi^2$. In fact, \cite{Deming_2011} explain the anomalous eclipse depth by invoking emission from a circumstellar accretion disk contaminating the system flux in the mid-infrared at the level of $5\times10^{-3}$.

\cite{Chen_2014} performed spectral retrieval on WASP-5b, but the authors were unable to obtain a good fit that conserved energy, even when they allowed the atmospheric C/O ratio to vary.  It is hard to imagine that a self-consistent radiative transfer model with only two variables would do any better.  

The 3.6~$\mu$m photometry of HD~189733b is $5\times10^{-4}$ discrepant from the best match 1D radiative transfer model obtained by varying two model parameters \citep[][]{Knutson_2012}. The mismatch between the predicted and measured flux at 3.6~$\mu$m contributes $(5\times10^{-4}/4\times10^{-5})^2 = 156$ to the $\chi_{\rm RT}^2$ budget, making this model a far worse fit than a simple blackbody ($\chi_{\rm BB}^2 = 33$, as shown in Figure~\ref{blackbody_BIC}).

It is likely that bona fide fits using self-consistent radiative transfer models could provide somewhat better $\chi_{\rm RT}^2$, but this is computationally intensive and has only been performed once, to our knowledge \citep{Kipping_2011}. A recent wholesale look at all extant eclipse spectra concluded that the only potentially robust area of agreement between self-consistent models and the data was the ``systematic increase in the ratios to shorter wavelengths'' \citep{Burrows_2014}.    

In other words, the planets poorly fit by blackbodies are also poorly fit by self-consistent radiative transfer models. The radiative transfer models could simply be wrong. There have been efforts to compare and validate exoplanet radiative transfer codes \citep{Guillot_2010, Shabram_2011} and many have been tested against high quality observations of brown dwarfs, but it is possible that they are missing important physics relevant to irradiated planets. ``Missing physics'' includes atmospheric dynamics and clouds, but these are also omitted from most spectral retrieval models.  We therefore hypothesize that the spectral features in extant broadband spectra are due to a combination of astrophysical + detector noise\footnote{The possibility that features in broadband hot Jupiter emission spectra are merely a combination of detector and astrophysical error has previously been noted by G.P.~Laughlin: http://oklo.org/2013/08/21/central-limit-theorem}; spectral retrieval provides better fits because it is under-constrained.

\subsection{Are New Measurements More Accurate?}
Most recent measurements have not yet been repeated, but one could argue that the various advances in reduction and analysis have made modern eclipse measurements more accurate than their predecessors.  In hindsight, it is easy to point out poor judgements made by earlier researchers. In all cases, however, the authors were making defensible choices about how to treat the data and how to fit it. In no case has the original paper been retracted or has an erratum been published.  With one exception \citep{Beaulieu_2011}, researchers have only questioned the original measurements once better observations were available.  

Researchers still make choices about their reduction scheme, and the intra-pixel sensitivity variations of \emph{Warm Spitzer} are still modeled using the same few hours of data that are used to measure the eclipse depth. We should aspire to parametrize these choices and marginalize over them to produce accurate, if less precise, measurements. A promising avenue is to use Gaussian Processes to model the intrapixel sensitivity variations, which implicitly marginalizes over the functional form of the detector model. This strategy has been used for transit spectroscopy \citep[][]{Gibson_2012, Gibson_2013} and to model the effect of star spots on thermal phase variations \citep{Knutson_2012}.

Moreover, none of the studies reporting secondary eclipse measurements account for how the meta-parameters of reduction and analysis pipelines contribute to uncertainty in eclipse depth.  At best, researchers experiment with a variety of schemes and adopt the one that minimizes the scatter in the photometry \citep[][]{Stevenson_2012}.  This amounts to optimizing the meta-parameters rather than marginalizing over them.  If different choices of meta-parameters, detector parametrization, or astrophysical parametrization lead to significantly different eclipse depths (see \S4.2), then one should be wary of small quoted uncertainties. 

The possibility of multimodal posterior distributions should also give us pause, since neither gradient descent (e.g., Levenberg-Marquardt) nor Markov Chain Monte Carlo routines are well suited to finding global solutions under these circumstances.

In short, the current generation of single-eclipse measurements are still systematics-dominated and susceptible to many of the same problems as the previous generation.  In the $\sigma_{\rm syst}$ hypothesis, there is a noise floor that affects all single-eclipse measurements, so current single-eclipse measurements are little better than the first generation.  In the $f_{\rm syst}$ hypothesis, on the other hand, the uncertainties are under-estimated by a constant \emph{factor}, so single-eclipse measurements published today (which tend to have small quoted uncertainties) are taken to be more accurate than their predecessors.  In other words, the $f_{\rm syst}$ hypothesis assumes that eclipse depth estimates are becoming more accurate with time.\footnote{It may eventually be possible to repeat this study but with so many measurements in Table~\ref{reshoots} that $f_{\rm syst}$ can be a function of time, rather than constant; one could hope that  $f_{\rm syst}$ tends to unity, indicating that observers are getting better at estimating the accuracy of their measurements.}  Our results are independent of which hypothesis we choose, as discussed above. 

\subsection{Astrophysical Sources of Error}
Measurement-to-measurement variance in eclipse depths is only sensitive to systematics that change from epoch to epoch: detector behavior, star spots, and exoplanet weather. There are other systematics, however, that might remain constant from epoch to epoch but that still introduce an error in our estimate of the planetary flux.  

WASP-12b is the poster-child for such astrophysical sources of uncertainty, starting with the possibility of contamination from a circumstellar disk \citep{Li_2010}. A change in astrophysical assumptions---namely the strength of ellipsoidal variations---affects the 4.5~$\mu$m eclipse depth of WASP-12b by $1.1\times10^{-3}$ \citep{Cowan_2012}.

Moreover, published eclipse measurements of WASP-12b have had to be revised after the discovery of a binary companion that diluted the eclipse measurements, leading to eclipse depth increases of $8\times10^{-5}$ to $6.5\times10^{-4}$ in the near to mid-infrared \citep[][]{Crossfield_2012}. In short, even if the photometry for an exoplanet system were precisely known, there is significant room for error in the dayside emission of the planet, which is the quantity we need to know for spectral fitting.

\section{Conclusions}
The retrieval of atmospheric structure and composition from disk-integrated broadband photometry hinges on planets not emitting like blackbodies. We have considered the 44 short-period planets with emission measurements in multiple broadbands.  If published uncertainties are taken at face value, then seven of these planets have broadband spectra that favor spectral retrieval over blackbody fits, according to the Bayesian Information Criterion---CoRoT-2b benefits the most from the additional model parameters.  

In order to perform under-constrained spectral retrieval, however, it is critical to know the actual uncertainty on eclipse measurements. \emph{Spitzer} is capable of exquisite photometry ($<10^{-4}$), but single eclipses acquired, reduced and analyzed in isolation have only been repeatable at the $1\sigma$ level of $5\times10^{-4}$ (or single-eclipse uncertainties have been under-estimated by a factor of 3). If one adopts such empirical uncertainties for single-eclipse measurements, then blackbody fits are preferable over spectral retrieval for all planets, with the possible exception of HD~189733b. 

We conclude that statements about atmospheric composition based solely on broadband emission measurements are premature.  If one adopts empirical estimate of single-eclipse accuracy, then HD~209458b and GJ~436b are well fit by blackbodies, and WASP-12b is not so poorly fit as to favor spectral retrieval. This resonates with the cautionary review of \cite{Burrows_2014}. Temperature inversions and odd compositions were inferred for short period planets based on broadband emission spectra \citep{Knutson_2008, Knutson_2010, Stevenson_2010, Madhusudhan_2011, Madhusudhan_2012}. Our results call these claims into question. Undoubtedly, many planets have stratospheric inversions and non-solar chemistry, but there is no robust evidence for this in the current photometry of short-period planets.

\section*{Acknowledgments}
CJH was funded by the Illinois Space Grant. JCS was funded by an NSF GK-12 fellowship. NBC is indebted to the participants of ExoPAG-9 for discussions of instrument systematics, as well as to M.R.~Line and N.~Madhusudhan for discussions of spectral retrieval.  The authors received useful feedback from S.J.~Carey, D.~Dragomir, H.A.~Knutson, K.B.~Stevenson, J.N.~Winn, and the anonymous MNRAS referee. NBC acknowledges the generous hospitality of the Institut the Plan\'etologie et d'Astrophysique de Grenoble (IPAG).  This research has made use of the Exoplanet Orbit Database and the Exoplanet Data Explorer at exoplanets.org.

\bsp

\label{lastpage}

\end{document}